\documentclass[a4paper,11pt]{article}
\usepackage{pos}

\newcommand\beq{ \begin{eqnarray} }
\newcommand\eeq{ \end{eqnarray} }

\usepackage{color}

\title{Flux tube profiles in two-color QCD at low temperature and high density}
\author*[a]{Katsuya Ishiguro}
\affiliation[a]{Library and Information Technology, Kochi University, Kochi 780-8520, Japan}
\emailAdd{ishiguro@kochi-u.ac.jp}

\author[b]{Kei Iida}
\affiliation[b]{Department of Mathematics and Physics, Kochi University, 2-5-1 Akebono-cho, Kochi 780-8520, Japan}
\emailAdd{iida@kochi-u.ac.jp}

\author[c,d,e,f]{Etsuko Itou}
\affiliation[c]{Strangeness Nuclear Physics Laboratory,
RIKEN Nishina Center, Wako 351-0198, Japan}
\affiliation[d]{Interdisciplinary Theoretical and Mathematical Sciences Program (iTHEMS), RIKEN, Wako 351-0198, Japan}
\affiliation[e]{Department of Physics, and Research and 
Education Center for Natural Sciences, Keio University, 4-1-1 Hiyoshi, Yokohama, Kanagawa 223-8521, Japan}
\affiliation[f]{Research Center for Nuclear Physics (RCNP), Osaka University, Osaka 567-0047, Japan}
\emailAdd{itou@yukawa.kyoto-u.ac.jp}

\abstract{We investigate the temperature and density dependence of the color flux tube structure of dense two-color QCD with $N_f=2$ Wilson fermions by using a lattice simulation. From Refs.~\cite{Iida:2021} and \cite{Iida:2020}, we have already clarified the rich phase structure in the low temperature region, including the hadronic and superfluid phases. In this study we measure the quark-antiquark potential and color flux tube profiles in such a low temperature region and find that even in the high density superfluid phase, the color electric field is squeezed into a flux tube as in the low density hadronic phase.}

\FullConference{%
 The 38th International Symposium on Lattice Field Theory, LATTICE2021
  26th-30th July, 2021
  Zoom/Gather@Massachusetts Institute of Technology
}

\begin{document}
\maketitle

%%%%%%%%%%%%%%%%%%%%%%%%%%%%%%%%%%%%%%%%%%
\section{Introduction}
%%%
It is widely believed that QCD at low temperature and high density 
exhibits superconductivity.
Such superconductivity could appear inside neutron stars.
However, to elucidate QCD in such an extreme situation from first-principles calculations is a hard task because of the infamous sign problem.

In Ref.\ \cite{Iida:2020}, authors of these proceedings have studied the phase diagram of two-color QCD even in such a low temperature and high density regime, which is basically consistent with several works by other collaborations~\cite{Cotter:2013,Boz:2020,Bornyakov:2018,Astrakhantsev:2020}.
In these works, to avoid the sign problem, two-color QCD is considered, and also to solve the numerical instability (onset problem), the diquark source term is introduced in the finite-density QCD action.
It is found that the expectation value of the diquark condensate takes nonzero values in the high density regime, which is a clear signal of superfluidity.

In these proceedings, we investigate the string tension, which tells us the confinement property, and the color flux tube profiles between a quark and anti-quark pair.
As for the confinement, we have studied the Polyakov loop and the topological charge in the previous studies~\cite{Iida:2021,Iida:2020}, and have found that the confinement remains even in the superfluid phase.
However, the other work~\cite{Bornyakov:2018} has shown that the string tension decreases with increasing density and the deconfinement phase transition occurs  at sufficiently high density. 
As for the color flux tube profiles, we examine the density dependence of color electric fields and estimate the Ginzburg-Landau (GL) parameter, which is a characteristic parameter of the dual superconductivity, from the color electric fields.

The organization of this report is as follows: In Section~\ref{section:setup} we briefly explain the numerical setup in this work. In Section~\ref{section:results}, the numerical results for the density dependence of the static potential and the color flux tube profiles between a quark and anti-quark pair. Finally our findings are summarised in Section~\ref{section:summary}.

\section{Numerical setup} \label{section:setup}

\subsection{Lattice action}

Here, we utilize the Iwasaki gauge action, which is composed of the plaquette term with $W^{1\times 1}_{\mu\nu}$ and the rectangular term with $W^{1\times 2}_{\mu\nu}$,
\beq
S_g = \beta \sum_x \left(
 c_0 \sum^{4}_{\substack{\mu<\nu \\ \mu,\nu=1}} W^{1\times 1}_{\mu\nu}(x) +
 c_1 \sum^{4}_{\substack{\mu\neq\nu \\ \mu,\nu=1}} W^{1\times 2}_{\mu\nu}(x) \right) ,
\eeq
where $\beta = 4/g^2_0$ in the two-color theory, and $g_0$ denotes the bare gauge coupling constant. The coefficients $c_0$ and $c_1$ are set to $c_1 = -0.331$ and $c_0 = 1 - 8c_1$.

As for a lattice fermion action, we use the two-flavor Wilson fermion action.
Here, we add the quark number operator in the QCD fermion action and  we also introduce the diquark source term to study the low temperature and high density regime.
The diqaurk source term, which explicitly breaks U(1) baryon symmetry, is required to avoid the numerical instability.
The total fermion action is represented by
\beq
S_F= \bar{\psi}_1 \Delta(\mu)\psi_1 + \bar{\psi}_2 \Delta(\mu) \psi_2 - J \bar{\psi}_1 (C \gamma_5) \tau_2 \bar{\psi}_2^{T} + \bar{J} \psi_2^T (C \gamma_5) \tau_2 \psi_1.\label{eq:action}
\eeq
Here, the indices $1$, $2$ denote the flavor label, and $\mu$ is the quark chemical potential.
The additional parameters $J$ and $\bar{J}$ correspond to the anti-diquark and diquark source
parameters, respectively. For simplicity, we put $J = \bar{J}$ and assume that it takes a real
value. The Wilson-Dirac operator including the number operator, $\Delta(\mu)$, is defined by
\beq 
\Delta(\mu)_{x,y} = \delta_{x,y} 
&-& \kappa \sum_{i=1}^3  \left[ ( 1 - \gamma_i)  U_{x,i}\delta_{x+\hat{i},y} + (1+\gamma_i)  U^\dagger_{y,i}\delta_{x-\hat{i},y}  \right] \nonumber\\ 
&-& \kappa   \left[ e^{+\mu}( 1 - \gamma_4)  U_{x,4}\delta_{x+\hat{4},y} + e^{-\mu}(1+\gamma_4)  U^\dagger_{y,4}\delta_{x-\hat{4},y}  \right],
\eeq 
where $\kappa$ is the hopping parameter.

\subsection{Simulation parameters}

We perform the simulations with $\beta=0.800$ and $\kappa=0.159$ on a $32^4$ lattice. These parameters give the pseudoscalar to vector meson mass ratio $m_{\text{PS}}/m_{\text{V}}=0.8232$ and the temperature $T=0.19T_c$~\cite{Iida:2021,Iida:2020}. 
Here, $T_c$ denotes the psuedo-critical temperature determined by the peak of the chiral susceptibility.
The number of configurations for each set of $\mu$ and $J$ is 200 in this work.

%%%%%%%%%%%%%%%%%%%%%
\begin{table}[h]
\begin{tabular}{c|c|c|c|c|c|c|c|c|c} \hline
$\mu$ & 0.00 & 0.10 & 0.25 & 0.27 & 0.30 & 0.35 & 0.40 & 0.50 & 0.70 \\ \hline
$j$   & 0.00 & 0.00 & 0.00 & 0.02 & 0.02 & 0.02 & 0.02 & 0.02 & 0.02 \\ \hline
Phase & Hadronic & Hadronic & Hadronic & Hadronic & Border & BEC & BEC & BCS & BCS \\ \hline
\end{tabular}
\caption{The adopted values of the chemical potential $\mu$ and the diquark source $j=J/\kappa$ in the lattice unit and their corresponding phases}
\label{Table:parameter}
\end{table}
%%%%%%%%%%%%%%%%%%%
The adopted values of chemical potential $\mu$ and the diquark source $j=J/\kappa$ in lattice unit are summarised in Table~\ref{Table:parameter}.
According to the previous work, we have found that there are hadronic, Bose-Einstein condensed (BEC) and BCS phases at low temperature.
Here, in the BEC and BCS phases, the expectation value of the diquark condensate ($\langle qq \rangle$), which is the order parameter of superfluidity, is non-zero in contrast to the hadronic phase where $\langle qq \rangle =0$.
Furthermore, the perturbative picture is reasonable in the BCS phase, where the quark number density is almost consistent with the ideal-gas analysis.
For all the values of $\mu$ adopted here, the expectation value of the Polyakov loop is almost zero,  which indicates that the system is confined even at the highest density.

\section{Numerical results} \label{section:results}

\subsection{The quark-antiquark potentials}

We begin by observing the static potential between a quark and anti-quark pair to see if confinement-deconfinement phase transition occurs in a regime of low temperature and high density. The static potential $V(r)$ can be calculated from the Wilson loop $W(r,t)$ with spatial separation $r$ and temporal extent $t$ as
\beq
aV(r) = \lim_{t \rightarrow \infty} \log \frac{\langle W(r,t) \rangle}{\langle W(r,t+1) \rangle} ,
\eeq
where $a$ denotes the lattice spacing. To improve the signal-to-noise ratio, the APE smearing~\cite{APE:1987} and the hypercubic blocking~\cite{Hasenfratz:2001} are applied to the spatial links and the temporal links, respectively. 
%%%%%%%%%
\begin{figure}[htbp]
\begin{minipage}[t]{0.49\linewidth}
\begin{center}
\includegraphics[scale=0.45]{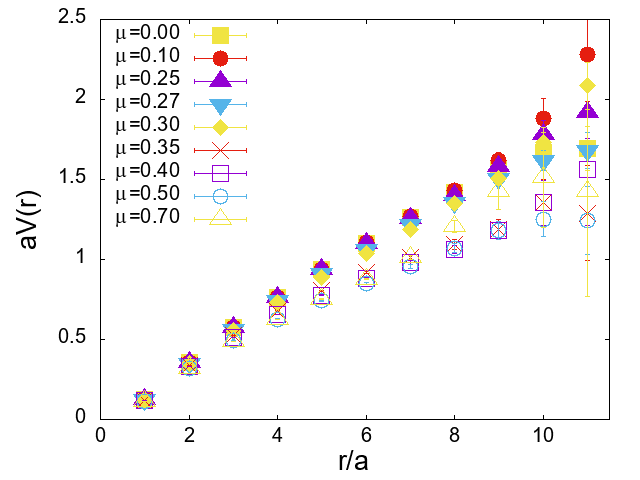}
%\caption{The static potentials and their $\mu$-dependence on $32^4$ lattices.}
%\label{fig:potential_mu}
\end{center}
\end{minipage}
%\end{figure}
%%%%%%%%%%%%%%
%%%%%%%%%
%\begin{figure}[htbp]
\begin{minipage}[t]{0.49\linewidth}
\begin{center}
\includegraphics[scale=0.45]{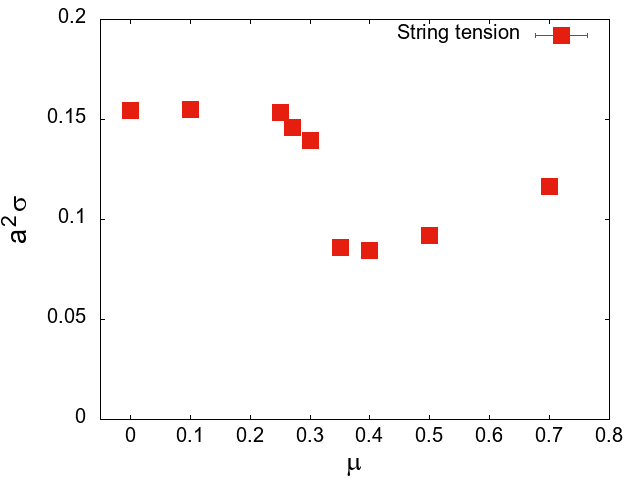}
%\caption{The $\mu$-dependence of the string tensions $\sigma$.}
%\label{fig:string_mu}
\end{center}
\end{minipage}
\caption{(Left) The $\mu$-dependence of the static potentials. The lattice size is $32^4$. (Right) The $\mu$-dependence of the string tensions ($\sigma$).}
\label{fig:potential_string_mu}
\end{figure}
%%%%%%%%%%%%%%

The $\mu$-dependence of the static potentials is shown in the left panel in Fig.~\ref{fig:potential_string_mu}. 
Although the static potentials have a weak $\mu$-dependence, the data of $aV(r)$ at large $r$ depict a linear behaviour for all the values of $\mu$.  Thus, the confinement potential emerges even at the highest density, which is consistent with the behavior of the Polyakov loop.

We also estimate the $\mu$-dependence of the string tension by fitting these potentials using the Cornell type function,
\beq
V(r)=\sigma r + \frac{c}{r}+V_0,
\eeq
where $\sigma$, $c$ and $V_0$ denote the string tension, the Coulombic coefficient, and a constant, respectively.
The obtained string tension as a function of $a\mu$ is represented in the right panel in Fig.~\ref{fig:potential_string_mu}.  The result for the string tension is almost constant in the hadronic phase ($\mu <0.3$), while it decreases around $\mu=0.30$ but still has a non-zero value even at the highest $\mu$. This indicates that two-color QCD at $T=0.19T_c$ is confined in both BEC and BCS phases.

\subsection{Flux tube profiles}

\begin{figure}[htbp]
\begin{minipage}[t]{0.49\linewidth}
\begin{center}
\includegraphics[scale=0.5]{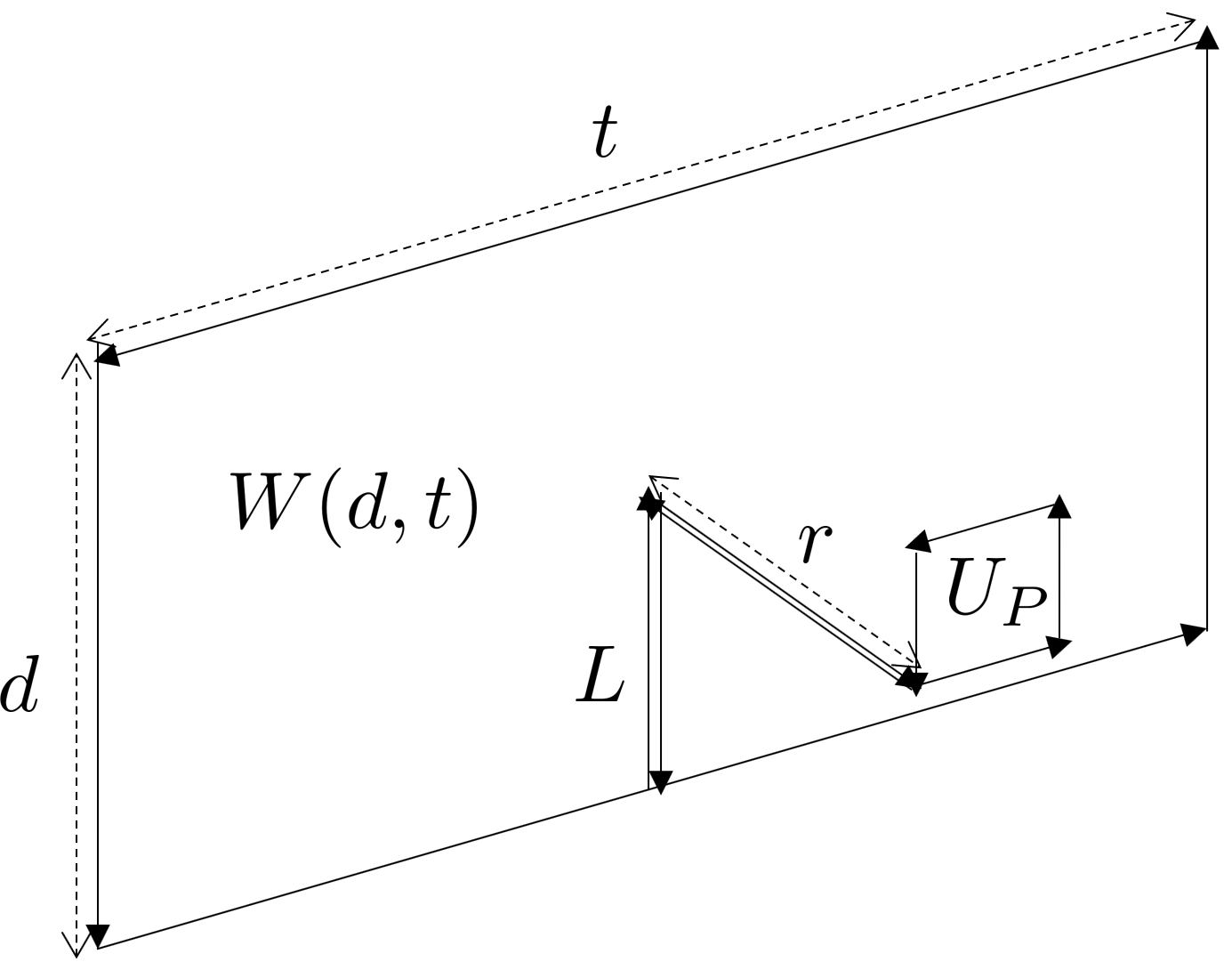}
%\caption{The connected correlator given in Eq.~(\ref{eq:connected}) between the Wilson loop $W$ and the plaquette $U_P$.}
%\label{fig:connected}
\end{center}
\end{minipage}
%\end{figure}
%\begin{figure}[htbp]
\begin{minipage}[t]{0.49\linewidth}
\begin{center}
\includegraphics[scale=0.6]{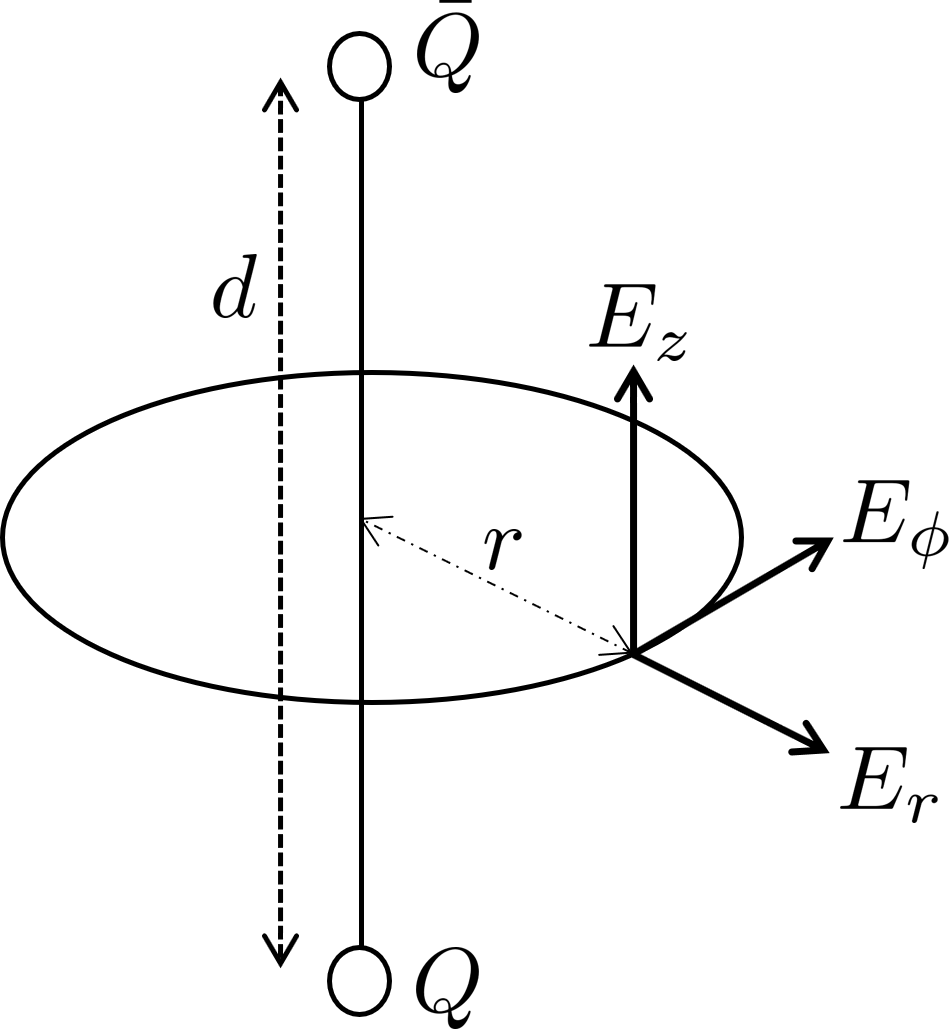}
%\caption{The color electric fields in cylindrical coordinate system. The d corresponds to distance between quark and anti-quark.}
%\label{fig:cylindrical}
\end{center}
\end{minipage}
\caption{(Left) The connected correlator given in Eq.~(\ref{eq:connected}) between the Wilson loop $W$ and the plaquette $U_P$ with the Schwinger line $L$. (Right) The color electric fields in cylindrical coordinate system. $d$ denotes the distance between a quark and an anti-quark.}
\label{fig:connected_cylindrical}
\end{figure}

Next, let us depict the flux tube profiles between a quark and an anti-quark to investigate the vacuum structure. To see the flux tube profiles, we consider the connected correlator $\rho_W$ between the Wilson loop $W$ and the plaquette $U_P=U_{\mu\nu}$ as done in Refs.~\cite{Cea:2017,Bonati:2018}:
\beq
\rho_W = \frac{\langle \mathrm{Tr}(WLU_PL^\dagger) \rangle}{\langle \mathrm{Tr}(W) \rangle}
-\frac{1}{N_c}\frac{\langle \mathrm{Tr}(U_P)\mathrm{Tr}(W) \rangle}{\langle \mathrm{Tr}(W) \rangle},
\label{eq:connected}
\eeq
where $N_c$ is the number of colors and $L$ is the Schwinger line (see the left panel in Fig.~\ref{fig:connected_cylindrical} for a pictorial representation). 
In the continuum limit, the connected correlator (\ref{eq:connected}) {\bf reduces} to $\rho_W \rightarrow a^2 g_0 \langle F_{\mu\nu} \rangle_{q\bar{q}}$, where the subscript $q\bar{q}$ is used to denote that a static quark and anti-quark pair is present in the background.
The APE smearing~\cite{APE:1987} and the hypercubic blocking~\cite{Hasenfratz:2001} are applied in the same way as the calculation of the static potentials.

\begin{figure}[htbp]
\begin{minipage}[c]{0.48\linewidth}
\centering
\includegraphics[scale=0.45]{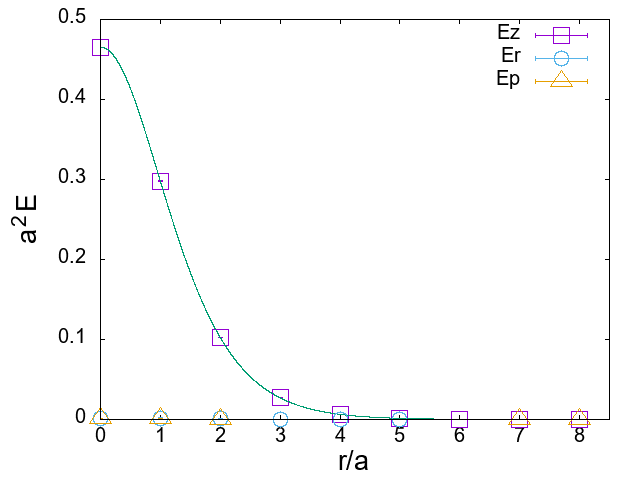}
%\caption{Electric fields at $\mu=0.00$.}
%\label{fig:Erpz_mu000}
\end{minipage}
\begin{minipage}[c]{0.48\linewidth}
\centering
\includegraphics[scale=0.45]{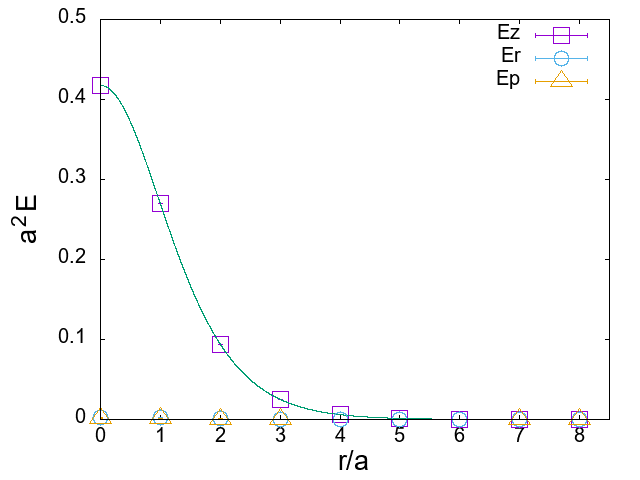}
%\caption{Electric fields at $\mu=0.50$.}
%\label{fig:Erpz_mu050}
\end{minipage}
\caption{The spatial distributions of color electric fields with $W(3,3)$ at $\mu=0.00$ (left) and $\mu=0.50$ (right). The solid line represents the fitted line by Eq.\ (\ref{eq:Clem01}). }
\label{fig:Erpz}
\end{figure}

\begin{figure}[htbp]
\begin{center}
\includegraphics[scale=0.5]{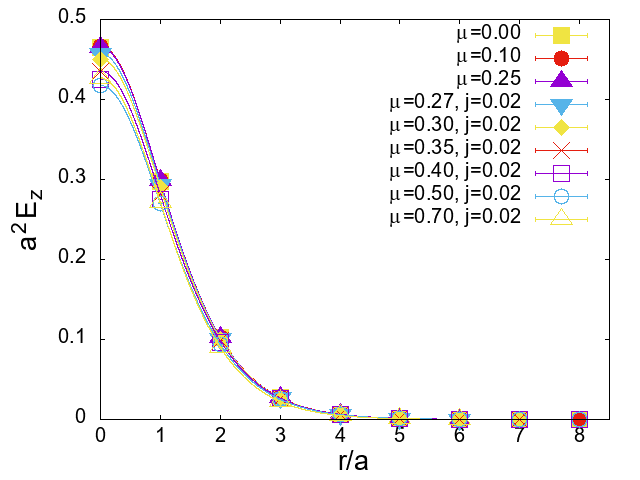}
\caption{The spatial distribution of the color electric fields $E_z$ and their $\mu$-dependence with $W(3,3)$.}
\label{fig:Ez_mu}
\end{center}
\end{figure}

We use the cylindrical coordinate $(r,\phi,z)$ to parametrize the quark and anti-quark system as shown in the right panel in  Fig.~\ref{fig:connected_cylindrical}.   
%%%%%%%%%
%%%%%%%%%%%%%%
The spatial distribution of color electric fields which is defined by the plaquette $U_{ti}$ with the Wilson loop $W(3,3)$ at $\mu=0.00$ (hadronic phase) and $\mu=0.50$ (BCS phase) are shown in Fig.~\ref{fig:Erpz}. 
Figure~\ref{fig:Ez_mu} shows the spatial distribution of color electric fileds $E_z$ and thier $\mu$-dependence. From Fig.~\ref{fig:Erpz} and Fig.~\ref{fig:Ez_mu}, we can see the radial and the azimuthal components of color electric fields are almost zero and only the $z$-components $E_z$ have non-zero values.  These results indicate that the color electric fields between a quark and an anti-quark are squeezed into tube-like structure in the range of $\mu =0.00$ to $\mu=0.70$. 

%%%%%%%%%
%%%%%%%%%%%%%%

%%%%%%%%%
%%%%%%%%%%%%%%

Assuming the dual superconductor picture of the QCD vacuum~\cite{tHooft:1976,Mandelstam:1976}, we fit the results for $E_z$ to the Clem form~\cite{Cea:2012,Clem:1975} as
\beq
E_z(r)=\frac{\phi}{2\pi\lambda^2\alpha}\frac{K_0[(r^2/\lambda^2+\alpha^2)^{\frac{1}{2}}]}{K_1[\alpha]} , \label{eq:Clem01} \\ 
\frac{1}{\alpha}=\frac{\lambda}{\xi_v} ,
\hspace{0.03\linewidth}
\kappa=\frac{\lambda}{\xi}=\frac{\sqrt{2}}{\alpha}\sqrt{1-\frac{K_0^2(\alpha)}{K_1^2(\alpha)}} ,
\label{eq:Clem02}
\eeq
where $\phi$, $\lambda$ and $\alpha$ are fit parameters and $K_0$, $K_1$ are the modified Bessel functions of the second kind. By fitting Eq.~(\ref{eq:Clem01}) to the data of $E_z$, we can estimate the penetration length $\lambda$, the variational core-radius $\xi_v$ and the coherence length $\xi$. The ratio of the penetration length and the coherence length $\kappa=\lambda/\xi$ is called the Ginzburg-Landau (GL) parameter, whose value discriminates between type I superconductors ($\kappa < 1/\sqrt{2}$) and type II superconductors ($\kappa >  1/\sqrt{2}$)~\cite{Tinkham:1996}.

%%%%%%%%%
\begin{figure}[htbp]
\begin{minipage}[c]{0.48\linewidth}
\begin{center}
\includegraphics[scale=0.45]{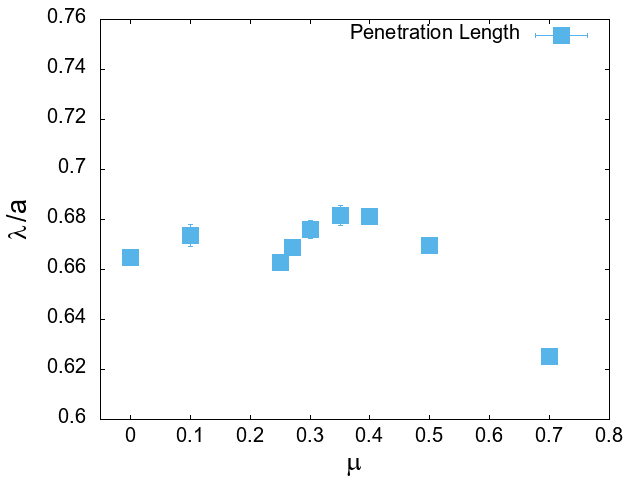}
%\caption{$\mu$-dependence of penetration length.}
%\label{fig:pene_mu}
\end{center}
\end{minipage}
%\end{figure}
%%%%%%%%%%%%%%
%%%%%%%%%
%\begin{figure}[htbp]
\begin{minipage}[c]{0.48\linewidth}
\begin{center}
\includegraphics[scale=0.45]{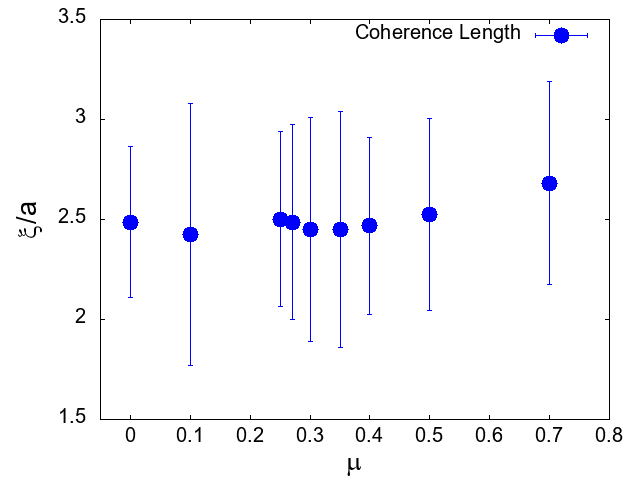}
%\caption{$\mu$-dependence of coherence length.}
%\label{fig:cohe_mu}
\end{center}
\end{minipage}
\caption{(Left) The $\mu$-dependence of the penetration length $\lambda$. (Right) The $\mu$-dependence of the coherence length $\xi$.}
\label{fig:pene_cohe_mu}
\end{figure}
%%%%%%%%%%%%%%

\begin{figure}[htbp]
\begin{center}
\includegraphics[scale=0.5]{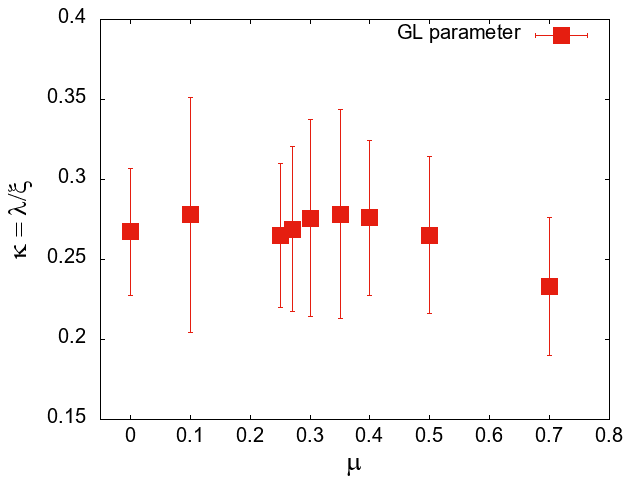}
\caption{The $\mu$-dependence of the Ginzburug-Landau parameter $\kappa = \lambda/\xi$.}
\label{fig:GL_mu}
\end{center}
\end{figure}

The left and right panels in Fig.~\ref{fig:pene_cohe_mu} present the $\mu$-dependence of the penetration length $\lambda$ and of the coherence length $\xi$, respectively. 
The penetration length appears to be larger near the border between the hadronic phase and the BEC phase, i.e., $0.30 \le \mu \le 0.40$, but the systematic error due to the small size of the Wilson loop should be considered for a final conclusion. On the other hand, $\lambda$ is significantly smaller in the high density region ($\mu=0.70$). The coherence length seems constant within error bars for all the values of $\mu$, although the error bars are large.  
The $\mu$-dependence of the GL parameter $\kappa$ estimated by the obtained results for the penetration and coherence lengths is shown in Fig.~\ref{fig:GL_mu}. $\kappa$ also seems constant within error bars for all the values of $\mu$, although the error bars are large due to the large error of $\xi$. Since the values of the GL parameter are less than $1/\sqrt{2}$, it is suggested that the vacuum of two-color QCD is a type I dual superconductor like ordinary metal superconductors.
%as seen in a single metal superconductors.  

%%%%%%%%%
%%%%%%%%%%%%%%

\section{Summary} \label{section:summary}

In this work, we investigate the density dependence of the static potentials and the color electric fields between a quark and anti-quark pair for $N_f=2$ two-color QCD with Wilson fermions at $T=0.19T_c$. It is found that the static potentials exhibit a linear behavior at large distance, in a manner depending only weakly on $\mu$. The result for the string tension varies around $\mu=0.30$ and has a non-zero value even at the highest $\mu$. 
The penetration length estimated by fitting to the color electric fields ($E_z$) has a peak around the critical point from the hadronic phase to the BEC phase, and is significantly small at high density.  
The coherence length and the GL parameter are almost constant within error bars at all $\mu$. Although it is needed to increase the number of statistics and to evaluate the systematic errors, the values of the GL parameter estimated are less than $1/\sqrt{2}$. It is suggested that the vacuum of two-color QCD is a type I dual superconductor. At $T=0.19T_c$, even at the highest density, the flux tube squeezing remains, and the static potential has nonzero string tension.
These results indicate that the system of $N_f=2$ two-color QCD at $T=0.19T_c$ is in a confined phase in the range of $\mu \le 0.70$ and are consistent with the previous results for the Polyakov loop and topological susceptibility~\cite{Iida:2021,Iida:2020}.

%%%%%%%%%%%%%%%%%%%%%%%
%%%%%%%%%%%%%%%%%%%%%%%
\subsection*{Acknowledgment}
%\begin{acknowledgments}
%%%%%%%%%%%%%%%%%%%%%%%
%%%%%%%%%%%%%%%%%%%%%%%
The work of E.~I. is supported by JSPS KAKENHI with Grant Numbers 
19K03875 and JP18H05407, JSPS Grant-in-Aid for Transformative Research Areas (A) JP21H05190,  JST PRESTO Grant Number JPMJPR2113 and the HPCI-JHPCN System Research Project (Project ID: jh210016).
The work of K.~I. is supported by JSPS KAHENHI with Grant Numbers 18H01211 and 18H05406.

%%%%%%%%%%%%%%%%%%%%%%%%%%%%%%%%%%%%%%%%%%%


\begin{thebibliography}{99}


%%\cite{Itou:2020azb}
%\bibitem{Itou:2020azb}
%E.~Itou and Y.~Nagai,
%%``Sparse modeling approach to obtaining the shear viscosity from smeared correlation functions,''
%JHEP \textbf{07} (2020), 007
%doi:10.1007/JHEP07(2020)007
%[arXiv:2004.02426 [hep-lat]].
%%2 citations counted in INSPIRE as of 20 Oct 2021


\bibitem{Iida:2021}
K.~Iida, E.~Itou and T.~G.~Lee, 
%“Relative scale setting for two-color QCD with $N_f=2$ Wilson fermions,” 
PTEP \textbf{2021} (2021), 013B05  
[arXiv:2008.06322 [hep-lat]].

\bibitem{Iida:2020}
K.~Iida, E.~Itou and T.~G.~Lee, 
%“Two-Colour QCD Phases and the Topology at Low Temperature and High Density,” 
JHEP \textbf{2001} (2020) 181 
[arXiv:1910.07872 [hep-lat]].

%\bibitem{Muroya:2001}
%S.~Muroya, A.~Nakamura, and C.~Nonaka, 
%[arXiv:hep-lat/0010073 [hep-lat]].

\bibitem{Cotter:2013}
S.~Cotter, P.~Giudice, S.~Hands and J.-I.~Skullerud, 
%"Towards the phase diagram of dense two-color matter,"
Phys. Rev. \textbf{D87}, 034507 (2013). 
[arXiv:1210.4496 [hep-lat]].

\bibitem{Boz:2020}
T.~Boz, P.~Giudice, S.~Hands and J.-I.~Skullerud, 
%"Dense two-color QCD towards continuum and chiral limits."
Phys. Rev. \textbf{D101}, 074506 (2020).
[arXiv:1912.10975 [hep-lat]].

\bibitem{Bornyakov:2018}
V.~G.~Bornyakov, V.~V.~Braguta, E.~M.~Ilgenfritz, A.~Y.~Kotov, A.~V.~Molochkov and
A.~A.~Nikolaev, 
%“Observation of Deconfinement in a Cold Dense Quark Medium,” 
JHEP \textbf{03} (2018), 161
[arXiv:1711.01869 [hep-lat]].

\bibitem{Astrakhantsev:2020}
N.~Astrakhantsev, V.~V.~Braguta, E.~M.~Ilgenfritz, A.~Yu.~Kotov and A.~A.~Nikolaev,
%“Lattice study of thermodynamic properties of dense QC2D,” 
Phys. Rev. \textbf{D102}, 074507 (2020), 
[arXiv:2007.07640 [hep-lat]].

\bibitem{APE:1987}
M.~Albanese et al. [APE Collaboration], 
Phys. Lett. \textbf{B192}, 163 (1987).

\bibitem{Hasenfratz:2001}
A.~Hasenfratz and F.~Knechtli, 
Phys. Rev. \textbf{D64}, 034504 (2001) 
[hep-lat/0103029].

\bibitem{Cea:2017}
%Flux tubes in the QCD vacuum
P.~Cea, L.~Cosmai, F.~Cuteri and A.~Papa,
Phys. Rev. \textbf{D95}, 114511 (2017) 
[arXiv:1702.06437 [hep-lat]].

\bibitem{Bonati:2018}
%Effects of a strong magnetic field on the QCD flux tube
C.~Bonati, S.~Calì, M.~D’Elia, M.~Mesiti, F.~Negro, A.~Rucci and F.~Sanfilippo, 
Phys. Rev. \textbf{D98}, 054501 (2018) 
[arXiv:1807.01673 [hep-lat]].

\bibitem{tHooft:1976}
G.~’t~Hooft, 
in Proceedings of the EPS International, edited by A. Zichichi, (1976), p.1225.

\bibitem{Mandelstam:1976}
S.~Mandelstam, 
Phys. Rep. \textbf{23}, 245 (1976).
 
\bibitem{Cea:2012}
%Chromoelectric flux tubes and coherence length in QCD
P.~Cea, L.~Cosmai and A.~Papa, Phys.Rev. \textbf{D86}, 054501 (2012) 
[arXiv:1208.1362 [hep-lat]].

\bibitem{Clem:1975}
J.~R.~Clem, Journal of Low Temperature Physics \textbf{18}, 427 (1975).

\bibitem{Tinkham:1996}
M.~Tinkham, “Introduction to superconductivity”, McGraw-Hill (1996).

\end{thebibliography}
\end{document}